\documentclass[reprint,nofootinbib,aps,prd,]{revtex4-2}

\usepackage{graphicx}
\usepackage{amsmath}
\usepackage{amssymb}
\usepackage{mathrsfs}
\usepackage{braket}
\usepackage{float}
\usepackage{bm}
\usepackage[colorlinks=true,citecolor=blue]{hyperref}

\newcommand{\be}{\begin{equation}}
\newcommand{\ee}{\end{equation}}
\newcommand{\bea}{\begin{eqnarray}}
\newcommand{\eea}{\end{eqnarray}}

\newcommand{\RomanNumeralCaps}[1]{\MakeUppercase{\romannumeral #1}}

\begin{document}

\title{Quantum wormholes at spatial infinity}

\date{}
\author{Jo\~{a}o Magueijo}
\email{j.magueijo@imperial.ac.uk}
\author{Ganga Singh Manchanda}
\email{ganga.manchanda20@imperial.ac.uk}
\affiliation{Abdus Salam Centre for Theoretical Physics, The Blackett Laboratory, Imperial College, Prince Consort Rd., London, SW7 2BZ, United Kingdom}

\begin{abstract}
We derive the interesting result that the two asymptotically flat Universes classically linked by the Einstein-Rosen bridge may also be quantum mechanically connected in their far out regions. 
This would be felt by the Newtonian potential far away from a black/white hole system, and raises the possibility of establishing communication via  perturbations. We obtain our results by means of wavepackets with a small variance in the mass, solving the equations derived from a maximally symmetry-reduced canonical quantisation method. Mass and a proxy of the Newtonian potential appear as canonical duals, leading to a Heisenberg uncertainty relation between the two. Coherent states are then built, which become non-semiclassical only in two regions: asymptotic spatial infinity (where unitarity forces the packets to ``feel'' the other asymptotic spatial infinity), and inside the horizon at $r=Gm$ where there is ringing. Whilst the latter has been noted in the literature, the former---the quantum wormhole at spatial infinity---seems to have eluded past scrutiny. Even under a coherent state there is a free parameter determining the distance beyond which the states becomes non-semiclassical, given the ambiguity in defining quadratures and squeezing, with departures eventually becoming of order one and cutting off the Newtonian potential. Further studies are required to examine the stability of these conclusions beyond their symmetry-reduced test tube.
\end{abstract}

\maketitle

\section{Introduction}

The difficulties in obtaining a complete quantum theory of gravity are immense. To make such a task less daunting, one often appeals to a systematic reduction of the number of degrees of freedom in a system, leading to a ``minisuperspace/midisuperspace'' approach. Whether one learns important lessons from these studies or dismisses them as acts of desperation, leading to artifacts not vindicated by the full theory, will not be discussed here. We will simply copy techniques which fared well in the context of homogeneous and isotropic Universes over to the alien field of spherically symmetric (and static) black holes. 

This is far from new and goes back to, at least, the work of~\cite{BCMN}. Since then many authors have attempted similar approaches, most notably~\cite{Kuchar}, whose comprehensive symmetry reduction set the bar for modern work. Their reduction derives from a foliation into hypersurfaces of constant ``time'' co-ordinate, $\Sigma_t$, just as in quantum cosmology. In the case of Schwarzschild spacetime, however, there is another option which takes maximal advantage of symmetry. One can choose a foliation in hypersurfaces of constant radius, $\Sigma_r$, with $r$ playing the role of ``time'' or lapse. Such a strategy has been adopted before in~\cite{York,Davidson,Boer}.
The technical advantage of using $r$ as ``time'' with foliations in $\Sigma_r$ is that the $\Sigma_r$ hypersurfaces are cylinders $\mathbb{R}\times S^2$ which possess a particularly simple induced 3D Ricci scalar. More recent works such as those by~\cite{Lucia,Lopez} focus on the important issues surrounding inner products and boundary conditions. 

The primary aim of this paper is to draw attention to a strange property of the ensuing quantum solutions. When one performs a 3+1 decomposition one finds a conserved momentum, $p_X$, which classically is related to the mass of the black hole. Its dual, $X$, can be seen as the evolution variable (the ``time'' canonically dual to the on-shell constant, in the sense of~\cite{JoaoLetter,JoaoPaper} or~\cite{Henneaux,Unruh,UnimodKuchar,UnimodLee1,Daughton,UnimodDaughton,Sorkin1,Sorkin2,Bombelli,UnimodLee2}). This $X$ turns out to be a variable which classically and in the weak field limit is proportional to the Newtonian potential $\Phi\sim-\frac{1}{r}$. This simple fact implies that the wavepackets (which have constant spread in $X$ and $m\sim p_X$) bloat out in $r$ as we go far away from the black hole, as implied by error propagation and the form of the function $X\sim-\frac{1}{r}$. This should affect the Newtonian potential as we move far away from the black hole. Furthermore, in the absence of finely tuned boundary conditions, unitarity forces the packets to cover the whole line $X\in (-\infty,\infty)$. Hence, as the wave function moves into $r\rightarrow\infty$ ($X\rightarrow 0^-$), the state becomes sensitive to the far out region of the space beyond the Einstein-Rosen bridge $r\rightarrow-\infty$ ($X\rightarrow 0^+$).

The physics is not dissimilar to that found in~\cite{Gielen}. Unitarity forces the wave function (say a coherent state in $X$) to become double peaked in $r$. 
The upshot is that we predict a cut off of the Newtonian force in the far out region of the black hole (on a scale determined by the spread, $\sigma_X$, of the wavepacket, which remains a free parameter for reasons explained later). This can be interpreted as the mirror region pulling matter out into the black hole from the other infinity. The effect is obviously purely quantum, so the above heuristic image should be taken with an appropriate grain of salt, but it does give a rough visual interpretation. This strange effect leaves the doors open for communication between the two regions to be possible via perturbations to this solution in far out regions. Further studies are required to examine the stability of these conclusions beyond their symmetry-reduced test tube.

The plan of this paper is as follows. 
In \hyperref[Sec:classical]{Section~\ref{Sec:classical}} we begin by describing a spherically symmetric and static spacetime in the Hamiltonian formalism under our ``traded'' foliation. We show how Hamilton's equations of motion imply that this spacetime must be Schwarzschild on-shell.
Then, in \hyperref[Sec:quantum]{Section~\ref{Sec:quantum}} we canonically quantise the reduced theory and solve in co-ordinate space for a wavefunction. We construct (WKB approximated) Gaussian wavepackets by summing over the mass of the black hole.
In \hyperref[wormhole]{Section~\ref{wormhole}} we tighten up issues of measure, probability and unitarity to prove the main result of this paper: the regions of spacetime, classically accessible only via the Einstein-Rosen bridge, are quantum mechanically connected at infinity and that this entails a suppression of the Newtonian force at finite distance. 

\section{Classical theory}\label{Sec:classical}

We start by laying out the classical theory which we will canonically quantise. Our work draws on~\cite{Kuchar,York,Davidson} with a few significant differences, which we will highlight as they appear. 

\subsection{Symmetry-reduced Hamiltonian}

Consider a spherically symmetric and static metric of  time-like$\backslash$space-like Lorentzian signature, under a foliation in $\Sigma_r$. This metric can be expressed in terms of a lapse function, $N(r)$, and two free functions, $X(r)$ and $Y(r)$ (our parametrisation leaves ambiguity in dimensions which we ignore for now):
\begin{align}
    ds_{\text{time-like}\backslash\text{space-like}}^2&=\mp e^{X}dt^2\pm N^2dr^2+e^{Y-X}d\Omega_2^2.
    \label{metric}
\end{align}
Birkhoff's theorem states that, on-shell and in vacuum, this metric must describe the spacetime exterior$\backslash$interior ($\text{e}\backslash\text{i}$) to a Schwarzschild black hole. Though this remains true even if the requirement of staticity is relaxed, as we have eliminated the shift, $N^t$, it is also consistent to eliminate $t$-dependence. Selecting a metric ansatz ``freezes'' degrees of freedom to establish the minisuperspace. It can, however, also fix gauge degrees of freedom at the action level. As discussed in~\cite{Motohashi}, this can be dangerous and we must ensure any fixing is complete. As demonstrated by~\cite{Davidson}, a complete fixing (one which eliminates the shift, $t$-dependence, and fixes the $S^2$ metric to be areal i.e. $e^{Y-X}=r^2$) introduces second class constraints upon quantisation and a ``time'' (i.e. $r$) dependence in their Hamiltonian. To avoid these, we keep the $S^2$ metric generic and treat the lapse function, $N$, as a gauge degree of freedom (only after varying the action).

By imposing our symmetries on the Einstein-Hilbert action, we are able to produce a Hamiltonian for the metric (\ref{metric}) (see \hyperref[app:one]{Appendix A} for derivation):
\begin{align}
    \mathcal{H}_{\text{e}\backslash\text{i}}=-2GNe^{\frac{X}{2}}\left[\pm e^{-Y}(p_X^2-p_Y^2)+\frac{1}{4G^2}\right],
    \label{hamiltonian}
\end{align}
where $G$ is Newton's gravitational constant. The canonical co-ordinate $N$ acts as a Lagrange multiplier to generate the Hamiltonian constraint $\mathcal{H}_{\text{e}\backslash\text{i}}\approx0$. For the remaining canonical pairs, $(X,p_X)$ and $(Y,p_Y)$, we find Hamilton's equations via the Poisson bracket, $g'=\{g,\mathcal{H}\}$:
\begin{align}
    X'_{\text{e}\backslash\text{i}}&=\mp4GNe^{\frac{X}{2}-Y}p_X,
    \label{X}
    \\
    Y'_{\text{e}\backslash\text{i}}&=\pm4GNe^{\frac{X}{2}-Y}p_Y,
    \label{Y}
    \\
    p_X'&=0,
    \label{pX}
    \\
    p_Y'&\approx\frac{Ne^{\frac{X}{2}}}{2G}.
    \label{pY}
\end{align} 
As a consistency check, dropping the shift ($N^t=0$) amounts to ignoring the momentum constraint ${\cal H}_t=0$, but the Dirac hypersurface deformation algebra still closes because trivially $\{{\cal H}(r),{\cal H}(r')\}=0$. Indeed the Hamiltonian contains no derivatives, so this is an ultra-local theory.

\subsection{$X$, $Y$ and the black hole}

We must relate our canonical co-ordinates, $X$ and $Y$, to the radial co-ordinate, $r$.
The equations (\ref{X}), (\ref{Y}), (\ref{pX}) and (\ref{pY}) can be directly solved (most easily in the gauge $Ne^{\frac{X}{2}}=1$) to produce a number of solutions which resemble Schwarzschild through various transformations. This is, however, unnecessary as Hamilton's equations must be equivalent to Einstein's equations such that any solution we obtain must be diffeomorphic to a sector of the Kruskal-Szekeres maximal extension. As such it suffices to show that, for both the exterior and interior, the metric (\ref{metric}) is Schwarzschild and Hamilton's equations are satisfied when
\begin{align}
    e^X&=\left|1-\frac{2Gm}{r}\right|,
    \label{eX}
    \\
    e^Y&=r^2\left|1-\frac{2Gm}{r}\right|,
    \\
    p_X&=-\frac{m}{2},
    \\
    p_Y&=\frac{r}{2G}-\frac{m}{2},
\end{align}
in the gauge $Ne^{\frac{X}{2}}=1$ and in units where $c=1$. We have thus shown that the metric is Schwarzschild on-shell and permits spherically symmetric (and static) fluctuations away from Schwarzschild off-shell. We also ascertain, from (\ref{pX}), that the black hole mass is a gauge-invariant constant of motion in phase space. 
\begin{figure}[H]
    \centering
    \includegraphics[scale=0.45]{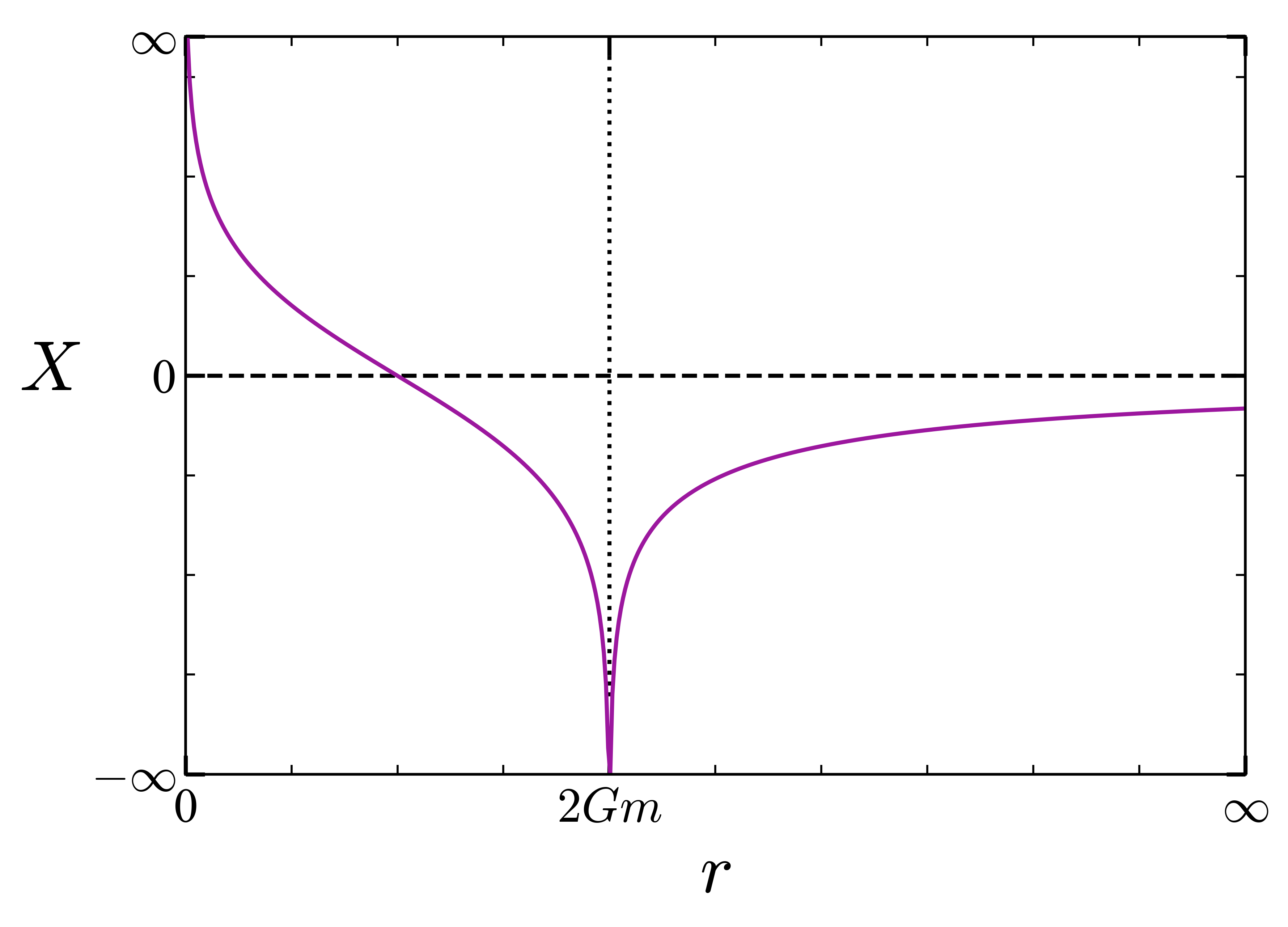}
    \caption{$r-X$ plot of the Schwarzschild solution for $r>0$. The $r=2Gm$ asymptote is drawn as a dotted line and the $X=0$ asymptote is drawn as a dashed line.}
    \label{rx}
\end{figure}

\hyperref[rx]{FIG.~\ref{rx}.} charts the relationship between $r$ and $X$, elucidating the connection to the on-shell black hole. The exterior range, $2Gm^+<r<\infty$, is mapped to $-\infty<X<0$ and the interior range, $0<r<2Gm^-$, is mapped to $-\infty<X<\infty$. 
\begin{figure}[H]
    \centering
    \includegraphics[scale=0.18]{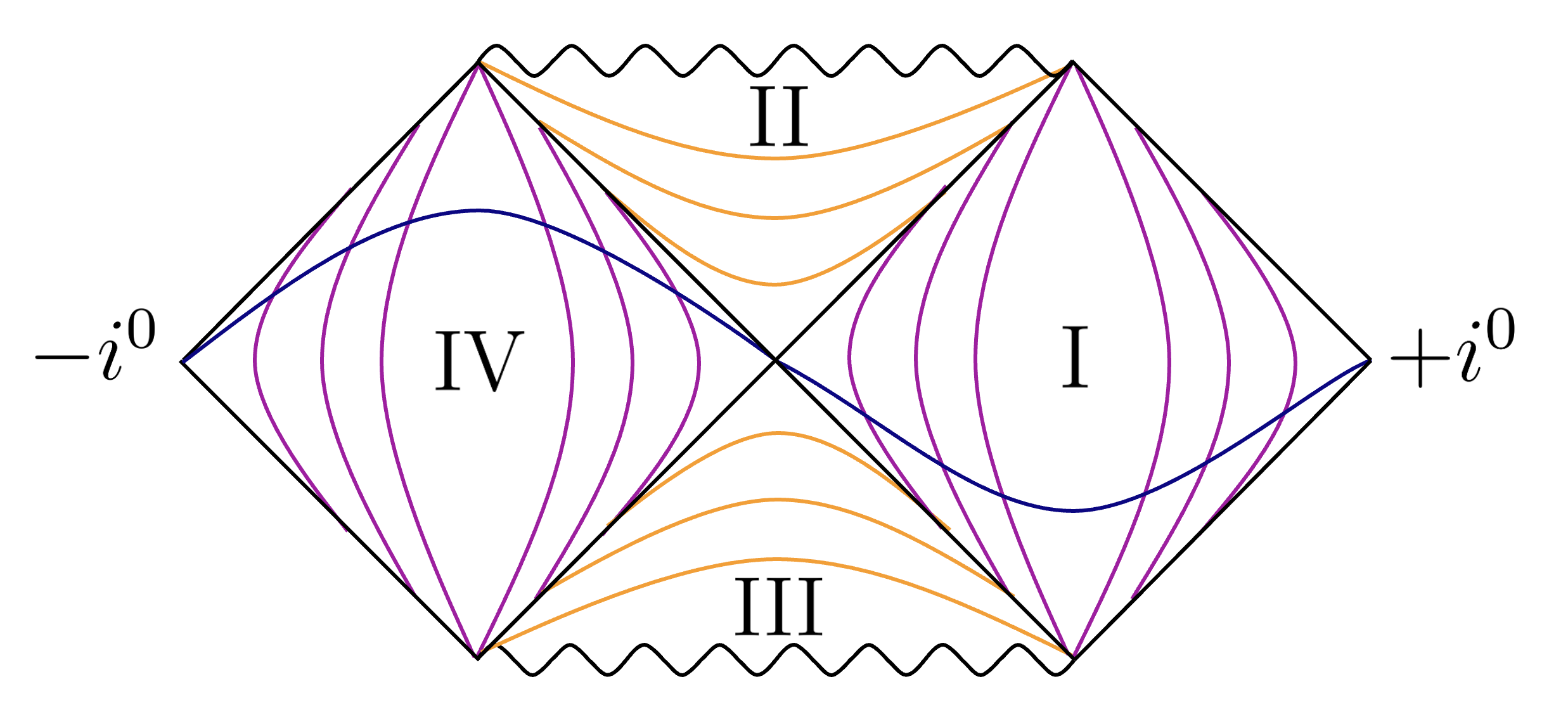}
    \caption{Carter-Penrose diagram divided into four regions: \RomanNumeralCaps{1}, \RomanNumeralCaps{2}, \RomanNumeralCaps{3}, and \RomanNumeralCaps{4}. Our foliation in surfaces of constant $r$ is drawn for the two separate cases of time-like $\Sigma_r$ (in purple, covering the exterior regions \RomanNumeralCaps{1} and \RomanNumeralCaps{4}) and space-like $\Sigma_r$ (in orange, covering the interior regions \RomanNumeralCaps{2} and \RomanNumeralCaps{3}). 
    A trajectory over the complete range of $X$ (see discussion in \hyperref[wormhole]{Section~\ref{wormhole}}) is drawn in blue between $-i^0$ and $+i^0$.}
    \label{penrose}
\end{figure}

\hyperref[penrose]{FIG.~\ref{penrose}.} outlines the complete spacetime allowing us to map the various regions to various ranges of $X$. Regions \RomanNumeralCaps{1} and \RomanNumeralCaps{4} are covered by two copies of $X\in(-\infty,0)$ and regions \RomanNumeralCaps{2} and \RomanNumeralCaps{3} are covered by two copies of $X\in(-\infty,\infty)$. Later we will cover the spacetime in an alternate atlas which we derive from unitarity. 

\begin{figure}[H]
    \centering
    \includegraphics[scale=0.45]{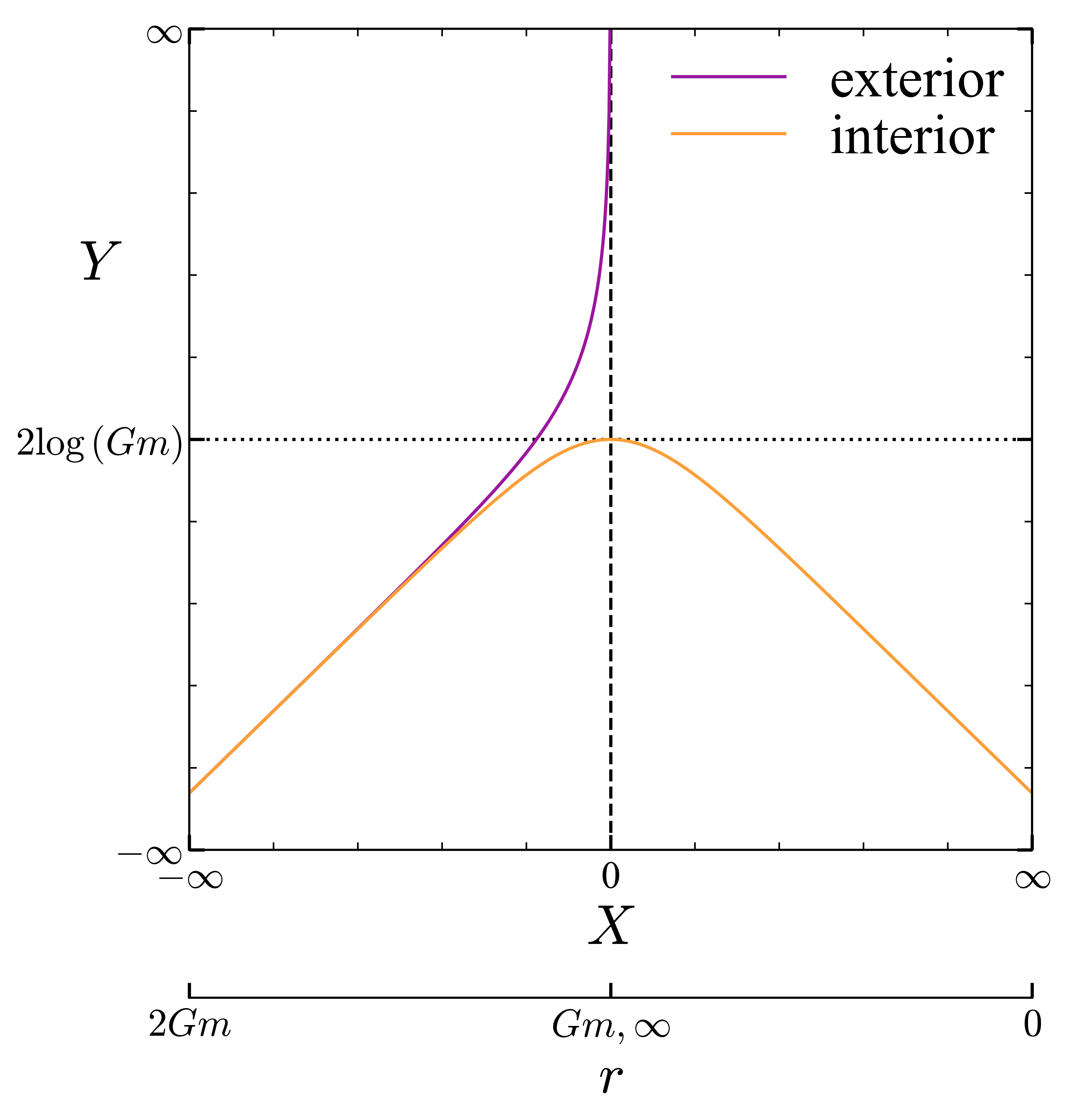}
    \caption{$X-Y$, plot of the exterior$\backslash$interior Schwarzschild trajectories. The corresponding $r$ values are also shown. The $X=0$ asymptote is drawn as a dashed line and $Y=2\log{(Gm)}$ is drawn as a dotted line.}
    \label{xy}
\end{figure}

Parametrically, we express the exterior$\backslash$interior Schwarzschild trajectories in configuration space as:
\begin{align}
    Y_{\text{e}\backslash\text{i}}=X+2\log{(2Gm)}-2\log{(1\mp e^X)}.
    \label{parametric}
\end{align}
\hyperref[xy]{FIG.~\ref{xy}.} graphs these trajectories and clearly demonstrates their linear behaviour in the $|X|\gg1$ limit.
This behaviour could also be inferred directly from Hamilton's equations which predict
\begin{equation}
    \frac{dY}{dX}=\frac{Y'}{X'}=\frac{r-Gm}{Gm}.
\end{equation}
The derivative makes four distinctions with respect to $r$:
\begin{equation} 
\frac{dY}{dX}\rightarrow
\begin{cases} 
    1,&\text{as }r\rightarrow 2Gm, \\
    \infty,&\text{as }r\rightarrow \infty, \\
    0,&\text{as }r\rightarrow Gm, \\
    -1, &\text{as }r\rightarrow 0.
\end{cases}
\end{equation}
Interestingly, the vanishing derivative at $r=Gm$ is indicative of a reflection or ``ringing'' even at the classical level though it is a purely quantum effect (see~\cite{Lopez} and also a similar effect in~\cite{Gielenringing,Alexandreringing}). A full discussion of the quantum properties of this effect would require studying the grey-body effect in black/white holes~\cite{grey1,grey2}, however the focus of our paper will not be on the physics near or inside the horizon. 

\section{Quantum theory}\label{Sec:quantum}

In this section, we canonically quantise to effectively reconstruct the ``fundamental'' quantum theory from its classical limit. However, the process comes with a catch: theories which are classically equivalent demonstrate no such equivalence once quantised. This makes quantisation a ``point of no return'' in the sense that we can flip variable ordering and trade foliation in the classical theory to no detriment, but once we quantise those choices are locked in and they undoubtedly impact the physics.

\subsection{Wheeler-DeWitt quantisation}

The constrained Hamiltonian, (\ref{hamiltonian}), can be canonically quantised to generate a Wheeler-DeWitt equation, $\hat{\mathcal{H}}_{\text{e}\backslash\text{i}}\ket{\Psi}=0$. As the equation possesses a global prefactor (under our choice of operator ordering), solutions need only satisfy
\begin{equation}
    \left(\hat{p}_X^2-\hat{p}_Y^2\pm\frac{e^{\hat{Y}}}{4G^2}\right)\ket{\Psi}=0.
\end{equation}
In a co-ordinate representation, the differential operator
\begin{equation}
    \hat{p}_q^2=-\hbar^2\frac{\partial^2}{\partial q^2},
\end{equation}
allows us to express the Wheeler-DeWitt equation as the hyperbolic partial differential equation:
\begin{align}
    \left(\frac{\partial^2}{\partial X^2}-\frac{\partial^2}{\partial Y^2}\mp\frac{e^Y}{4\hbar^2G^2}\right)\Psi(X,Y)=0.
    \label{wdw}
\end{align}
We note that the Planck length appears here as $\hbar^2G^2=l_P^4$. By taking a separable ansatz, $\Psi(X,Y)=\chi(X)\gamma(Y)$, with negative separation constant, $-\lambda^2$, we split (\ref{wdw}) into
\begin{gather}
    \frac{d^2\chi_\lambda}{dX^2}+\lambda^2\chi_\lambda=0,
    \\
    \frac{d^2\gamma_\lambda}{dY^2}+\left(\lambda^2\pm\frac{e^Y}{4l_P^4}\right)\gamma_\lambda=0.
    \label{gamma}
\end{gather}
Our choice of negative separation constant ensures oscillatory behaviour of $\chi_\lambda$:
\begin{equation}
    \chi_\lambda=c_+e^{i\lambda X}+c_-e^{-i\lambda X},
\end{equation}
which allows for an interpretation of $X$ as a physical clock with frequency, $\lambda$. Next, we note that under $\mathcal{Y}^2=\frac{e^Y}{l_P^4}$, (\ref{gamma}) can be written as the Bessel$\backslash$Modified Bessel equation:
\begin{equation}
    \mathcal{Y}^2\frac{d^2\gamma_\lambda}{d\mathcal{Y}^2}+ \mathcal{Y}\frac{d\gamma_\lambda}{d\mathcal{Y}}\pm\left[ \mathcal{Y}^2\mp(2i\lambda)^2\right]\gamma_\lambda=0.
\end{equation}
In the exterior, this is solved as
\begin{equation}
    \gamma_\lambda=k_+
    J_{2i\lambda}\left(\frac{e^{\frac{Y}{2}}}{l_P^2}\right)+k_-J_{-2i\lambda}\left(\frac{e^{\frac{Y}{2}}}{l_P^2}\right),
\end{equation}
while in the interior, we solve as~\cite{Gradshteyn}
\begin{equation}
    \gamma_\lambda=k_+e^{-\pi\lambda}I_{2i\lambda}\left(\frac{e^{\frac{Y}{2}}}{l_P^2}\right)+k_-e^{\pi\lambda}I_{-2i\lambda}\left(\frac{e^{\frac{Y}{2}}}{l_P^2}\right).
\end{equation}

\subsection{Exact Gaussian wavepackets}

Following~\cite{Kiefer,Lopez} (and paralleling~\cite{Alexandre,Gielen}), we build coherent wavepacket states by integrating the $\lambda$-mode wavefunctions, $\Psi_\lambda=\chi_\lambda\gamma_\lambda$, over an amplitude $\mathcal{A}\left(\lambda;\lambda_0,\sigma_\lambda\right)$. This is important because $\lambda$ is not some arbitrary quantum parameter, it carries physical meaning. As the clock frequency, $\lambda$, is conjugate to $X$ it is appropriate to treat it as the Fourier dual, $p_X=-\frac{m}{2}$. Such an interpretation leads to a generic black hole state which is a superposition over different masses:
\begin{equation}\label{packets2}
\begin{split}
    \Psi(X,Y;m_0,\sigma_m)=\int_\mathbb{R}\frac{dm}{2}\,&{\cal A}\left(-\frac{m}{2};-\frac{m_0}{2},\frac{\sigma_m}{2}\right)\cdot
    \\
    &\cdot\Psi_{-\frac{m}{2}}(X,Y).
\end{split}
\end{equation}
In this way we can characterise the black hole solely by mass related quantities, $m_0$ and $\sigma_m$, in agreement with the No-Hair theorem. To ensure a Gaussian distributed inner product we take an amplitude $\mathcal{A}=\sqrt{\mathcal{G}}$, where $\cal G$ is a Gaussian distribution. These wavepackets saturate the Heisenberg uncertainty relation, $\sigma_X\sigma_m\ge\hbar$. Taking the simplest boundary conditions which ensure an equal mixture of ingoing and outgoing waves, $c_+=c_-=k_+=k_-=1$, we write the exterior solution as
\begin{equation}
\begin{split}
    \Psi=&\left(\frac{2}{\pi\sigma_m^2}\right)^{\frac{1}{4}}\int_\mathbb{R}dm\,e^{-\frac{(m-m_0)^2}{4\sigma_m^2}}\cos{\left(\frac{mX}{2}\right)}\cdot
    \\
    &\cdot\left[J_{im}\left(\frac{e^{\frac{Y}{2}}}{l_P^2}\right)+J_{-im}\left(\frac{e^{\frac{Y}{2}}}{l_P^2}\right)\right],
    \label{tpacket}
\end{split}
\end{equation}
and the interior solution as
\begin{equation}
\begin{split}
    \Psi=&\left(\frac{2}{\pi\sigma_m^2}\right)^{\frac{1}{4}}\int_\mathbb{R}dm\,e^{-\frac{(m-m_0)^2}{4\sigma_m^2}}\cos{\left(\frac{mX}{2}\right)}\cdot
    \\
    &\cdot\left[e^{-\frac{m\pi}{2}}I_{im}\left(\frac{e^{\frac{Y}{2}}}{l_P^2}\right)+e^{\frac{m\pi}{2}}I_{-im}\left(\frac{e^{\frac{Y}{2}}}{l_P^2}\right)\right].
    \label{spacket}
\end{split}
\end{equation}

We naively expect $\mathcal{P}=|\Psi|^2$ to exhibit a ridge in configuration space which follows the classical trajectory for $m_0$ in a region of semiclassical correspondence. Unfortunately, without imposing non-generic boundary conditions not only can we not perform the integrals, numerical analysis is also unreliable due to divergences. Fortunately, the quantum cosmology literature provides a wealth of approximation know-how.

\subsection{Approximate Gaussian wavepackets}

We follow the WKB approximation scheme (see, e.g.~\cite{Bender}). To leading order (beyond eikonal order) this generates a wavefunction:
\begin{equation}\label{wkbpacket}
\begin{split}
    \Psi_{-\frac{m}{2}}^{\text{e}\backslash\text{i}}=&\left(-\frac{4}{l_P^4m^2\pm e^Y}\right)^{\frac{1}{4}}\left[c_+e^{-\frac{imX}{2}}+c_-e^{\frac{imX}{2}}\right]\cdot\\
    &\cdot\left[k_+e^{\frac{iP_{\text{e}\backslash\text{i}}}{l_P^2}}+k_-e^{-\frac{iP_{\text{e}\backslash\text{i}}}{l_P^2}}\right],
\end{split}
\end{equation}
where
\begin{equation}
\begin{split}
    P_{\text{e}\backslash\text{i}}(Y,m)=&\sqrt{l_P^4m^2\pm e^Y}
    \\
    &+\frac{l_P^2m}{2}\log{\left|\frac{\sqrt{l_P^4m^2\pm e^Y}-l_P^2m}{\sqrt{l_P^4m^2\pm e^Y}+l_P^2m}\right|}.
\end{split}
\end{equation}
This solution is complete in the sense that it possesses both ingoing and outgoing modes. Our earlier boundary conditions, $c_+=c_-=k_+=k_-=1$, impose an equal mixture such that these modes interfere to produce ``ringing'' effects at $r=Gm$. This phenomenon, and the interior as a whole, is studied by~\cite{Lopez} and interpreted as an annihilation process. We similarly produce an interfering wavepacket in the interior and extend the result to the exterior (see \hyperref[app:two]{Appendix B}), however as the literature has explored the former, we focus our attention on the latter. We expect no interference in the exterior such that it is sufficient to model a single outgoing mode under $c_+=k_+=1$, $c_-=k_-=0$. By inserting into \eqref{packets2}, expanding $P$ in Taylor series, and performing the Gaussian integral, we find the exterior solution:
\begin{equation}
\begin{split}
    \Psi=&\left[(-1)^{\frac{1}{4}}e^{i\left(\frac{P(m_0)}{l_P^2}-\frac{m_0X}{2}\right)}\right]\cdot
    \\
    &\cdot\left(\frac{8\pi\sigma_m^2}{l_P^4m_0^2+e^Y}\right)^{\frac{1}{4}}e^{-\frac{\sigma_m^2}{4}\left(X-Y_{\text{eff}}\right)^2},
    \label{wkb}
\end{split}
\end{equation}
with
\begin{equation}
\begin{split}
    Y_{\text{eff}}^{\text{e}\backslash\text{i}}(Y;m_0)&=\frac{2}{l_P^2}\left.\frac{\partial P_{\text{e}\backslash\text{i}}}{\partial m}\right|_{m_0}
    \\
    &=\log{\left|\frac{\sqrt{l_P^4m_0^2\pm e^Y}-l_P^2m_0}{\sqrt{l_P^4m_0^2\pm e^Y}+l_P^2m_0}\right|}.
\end{split}
\end{equation}
The peak of the Gaussian factor follows the trajectory
\begin{equation}
    X-Y_{\text{eff}}=0,
\end{equation}
whose positive branch can be written in the form of (\ref{parametric}) with $l_P^2$ taking the place of $G$,
\begin{equation}
    Y=X+2\log{(2l_P^2m_0)}-2\log{(1-e^X)}.
    \label{classical}
\end{equation}
This is nothing but the classical trajectory, as expected from a WKB solution. 

%
%
\section{The Quantum wormhole}\label{wormhole}

We now move on to the central part of our paper: the prediction of a quantum wormhole in the ``antipodes'' of the Einstein-Rosen bridge. This derives directly from unitarity. 

\subsection{Unitarity and the inner product}

Defining an inner product leading to unitarity can be problematic in these theories, and stumbles upon issues of boundary conditions (see for example the equivalent problems in quantum cosmology described in~\cite{GielenMenendez1,GielenMenendez2}). However, all difficulties evaporate if the operator associated with the conserved mass (i.e. $p_X$) can generate unrestricted translations in its dual variable (see for example the equivalent discussion in~\cite{Alexandre,Gielen} tracing back to~\cite{Isham}). Imposing unitarity in our context means forcing the wavepacket to propagate over the entire range $X\in(-\infty,\infty)$. As a result, rather than cover the spacetime with the atlas of copies described in \hyperref[Sec:classical]{Section~\ref{Sec:classical}}, we cover the spacetime with a single map between $-\infty<r<\infty$ and $-\infty<X<\infty$. The negative range of $r$ represents regions \RomanNumeralCaps{3} and \RomanNumeralCaps{4}, typically accessible only via the Einstein-Rosen bridge, through the transformation $r\rightarrow-r$. Under this mapping, a trajectory of the wavepacket peak in $X$ would appear as the blue line in \hyperref[penrose]{FIG.~\ref{penrose}.} moving through $+i^0$ and coming out on the other side at $-i^0$. Unitarity therefore enforces the existence of a new kind of wormhole.

The Hilbert space and inner product can be defined in analogy with what is done in unimodular-like theories\footnote{With the obvious replacements $\Lambda\rightarrow m$ (our black hole mass is like $\Lambda$, a classical constant of motion), $T_\Lambda\rightarrow X$ (our $X$ variable is like unimodular or 4-volume time), and $X_{\rm CS}\rightarrow Y_{\rm eff}$ (our $Y_{\rm eff}$ is like the Chern-Simons functional).}. The inner product is then defined as:
\be\label{innalpha}
\braket{\Psi_1|\Psi_2}=\int_\mathbb{R} \frac{dm}{2} \,  {\cal A}_1^\star\left(-\frac{m}{2}\right) {\cal A}_2\left(-\frac{m}{2}\right).
\ee
Such an inner product is automatically conserved with respect to the evolution variable $X$ (i.e. unitarity is enforced) since it is defined in terms of $X$-independent amplitudes. As explained in~\cite{Gielenringing} (see Sec.\RomanNumeralCaps{6}, in particular) we can now use Parseval's theorem to rewrite this inner product in terms of $Y_{\rm eff}$. In the case where the monochromatic waves (i.e. fixed $m$ solutions) are not plane waves, we linearize. This amounts to reducing to the eikonal approximation and writing our solutions as
\begin{equation}
\begin{split}
    \Psi =&\, e^{i\left(\frac{P(m_0)}{l_P^2}-\frac{m_0Y_{\text{eff}}}{2}\right)}\cdot
    \\
    &\cdot\int_\mathbb{R} \frac{dm}{2} \,  {\cal A}\left(-\frac{m}{2}\right)e^{-\frac{im}{2}(X-Y_{\text{eff}})}.
\end{split}
\end{equation}
Then we find the ``time''/$X$-independent inner product:
\be\label{innX}
\braket{\Psi_1|\Psi_2}=\int_\mathbb{R} dY_{\text{eff}} \, \Psi_1^\star(X,Y_{\text{eff}})\Psi_2(X,Y_{\text{eff}}).
\ee
This is an approximate inner product valid if we restrict ourselves to reasonably sharply peaked Gaussian wavepackets, centred on the same $m_0$ and obtained from the eikonal approximation. In this regime, a probability interpretation:
\begin{equation}\label{probdens}
\begin{split}
{\cal P}(Y_{\rm eff},X)&=\left|
\Psi(X,Y_{\text{eff})}\right|^2,
\end{split}
\end{equation}
is implied, where the density is in $Y_{\rm eff}$. 
By dialling $X$ we probe the probability distribution at regions with different values of $X\equiv X^{\text{classical}}=Y^{\text{classical}}_{\rm eff}$. 



\subsection{Corrections to the far out Newtonian potential}

One might wonder whether such a wormhole is traversable by means of quantum teleportation, for example. On a more mundane level the effect just found produces corrections to the far out Newtonian potential, $\Phi$, as we now show. This may be defined in the weak field limit from $Y_{\rm eff}\approx \pm2\Phi$, the sign depending on $\pm i^0$. The far out potential felt by classical bodies may be obtained from the smearing:
\begin{equation}
\begin{split}
     \overline{\Phi}&=\int_{\pm i_0} dr\,\mathcal{P}(r)\Phi\\
     &=\frac{1}{2}\left(\int_{-\infty}^0 dY_{\text{eff}}\, \mathcal{P}
     (Y_{\text{eff}},X) Y_{\text{eff}}\right.
     \\
     &\hspace{3em}\left.- \int_0^{\infty} dY_{\text{eff}}\, \mathcal{P}(Y_{\text{eff}},X) Y_{\text{eff}} \right),
\end{split}
\end{equation}
of packets peaked far out (on either region), that is for $X$ close to zero. For example, coherent states centred on the classical weak field result are described by a distribution, $\mathcal{P}(Y_{\text{eff}},X)=\mathcal{G}(Y_{\text{eff}};X\approx-\frac{Gm}{r},\sigma_X)$, such that:
\begin{equation}
     \overline{\Phi}=-\frac{Gm}{r}{\rm erf}\left(-\frac{\sqrt{2}Gm}{r\sigma_X}\right)+\frac{\sigma_X}{\sqrt{2\pi}}e^{\frac{-2G^2m^2}{r^2\sigma_X^2}}.
\end{equation}
\begin{figure}[H]
    \centering
    \includegraphics[scale=0.45]{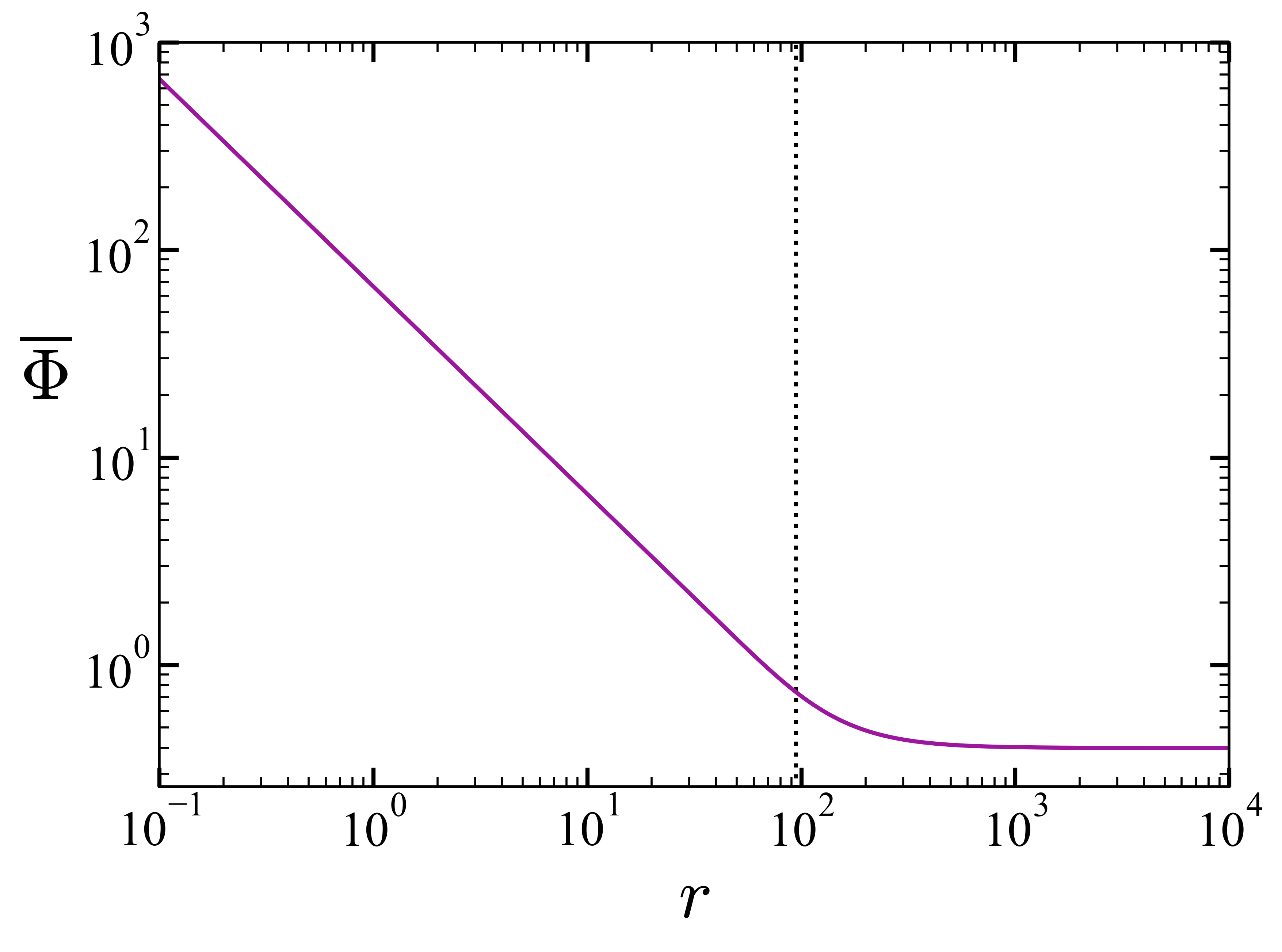}
    \caption{$r-\overline{\Phi}$, $\log-\log$ plot of the smeared Newtonian potential for $m=10^{12}$ and $\sigma_X=1$. $r_{\text{cut}}$ is drawn as a dotted line.}
    \label{potential}
\end{figure}
\hyperref[potential]{FIG.~\ref{potential}.} demonstrates how the potential converges to a non-zero constant, cutting off the Newtonian force, at a finite distance:
\begin{equation}\label{cutoff}
    r_{cut}\sim \frac{\sqrt{2} Gm}{\sigma_X}.
\end{equation}
Taking the most naive definition of dimensionless quadrature (and so of a coherent state) would imply $\sigma_X\sim 1$ and so a nonsensical cut off at the Schwarzschild radius. However the definition of squeezing is notoriously arbitrary in quantum gravity, so we should leave $\sigma_X$ arbitrary, and examine the phenomenology in those terms.

Note that another dimensionful scale in the problem is the mass itself (or the central mass around which the packet is built), such that we can redefine quadratures and instead write
\begin{equation}\label{cutoff1}
   r_{cut}\sim \frac{Gm}{l_P}\cdot\frac{\sigma_m}{m}.
\end{equation}
The last factor must obviously be less than one for the discussion to make sense (a matter that overlaps tangentially with the grey-body issues discussed before~\cite{grey1,grey2}). This provides a potential upper bound on $r_{cut}$, but it is not very useful. For a black/white hole with the mass of the Sun, our upper bound gives $r_{cut}=10^{38}~\mathrm{km}$, a distance much greater than the size of the Universe. 

Any phenomenology is therefore very much dependent on the free parameter, $\sigma_X$, which determines the distance from the black hole where departures from the classical behaviour become of order one. This in turn affects where the wavefunction becomes double peaked and where the Newtonian force is cut off. 

\section{Conclusions}

The study of ``quantum'' wormholes has a long history (see, for example,~\cite{Visser} and references therein, as well as the quantum sections of~\cite{Kuchar}). Here we picked up on some recent work on the subject~\cite{Lopez,Lucia,Davidson}, reproducing and agreeing with their conclusions, but unearthing a novelty that seems to have hitherto passed unnoticed. A minisuperspace approach can be devised, covering separately the interior and exterior regions, with time-like/space-like foliations (see \hyperref[penrose]{FIG.~\ref{penrose}.}). The mass appears as a conserved momentum conjugate to a variable, $X$, which may thus be employed as an evolution parameter. Wavepackets can be built, saturating the Heisenberg uncertainty relation. These behave semiclassically except in two regimes. One was noted before by~\cite{Lopez}, and results from a reflection leading to ``ringing'' (interference between the incident and the reflected packet); this occurs inside the black hole at $r=Gm$. The other regime is at asymptotic infinity,
linking the two asymptotically flat regions (usually only connected at $r=0$) at their ``end of the world'',   $\pm i^0$. This is a purely ``quantum wormhole'' (as opposed to the study~\cite{Visser}, where the same terminology is used).  

One may speculate about quantum contact between these regions, but interestingly we find a more prosaic effect: the cut off of the Newtonian force at a finite distance from the black hole. Heuristically, we can suggest that when the wavepacket starts to spread over the two asymptotic infinities, $\pm i^0$, we feel both the pull towards the black hole in the asymptotically flat region we are in (attached to, say, $+i^0$), and in the other regions (connected to $-i^0$), which translates as a pull {\it away} from the black hole (or an ``attraction towards infinity''). When the two exactly balance, for $r>r_{cut}$ (cf. \eqref{cutoff}), the force vanishes.  
We may then consider perturbations on top of this solution and speculate that contact could be classically established. Regardless of such speculation, we have the take-home fact that a cut off, not dissimilar to a Yukawa-like effect, is predicted. Obviously in our case the effect (and in contrast with Yukawa potentials) is purely quantum, and different systems could have different cut off scales, associated with different $\sigma_X$, as in \eqref{cutoff}. 


We close with a few general technical comments. 
We assumed Birkhoff's theorem before quantisation, which leads us back to the critical view in the opening paragraph: to what extent does symmetry reduction before quantisation lead to deceiving results and would a quantum analogue of Birkhoff's theorem resolve this issue? 
We defer this investigation to a future paper. On a more philosophical level (cf.~\cite{Isham}) it is curious that our pragmatic framework introduces a reverse ``block Universe'' problem. Usually we have a problem in that we treat time as space, thereby losing the sense of flow of time. Here we used a proxy for the radius as an evolution parameter in a framework that ultimately replaces the timeless Wheeler-DeWitt equation by a Schr\"{o}dinger-like equation (or Klein-Gordon-like). In a sense we introduce a ``flow of space''. Does this add to (or subtract from) the problem of time in quantum gravity?

\section{Acknowledgements}

We thank Andrew Tolley and Paolo M. Bassani for helpful discussions, and Aharon Davidson for early correspondence regarding this project. The work of JM was partly supported by the STFC Consolidated Grant ST/X00575/1.

\appendix

\section{Hamiltonian derivation}
\label{app:one}

We derive the exterior Hamiltonian starting with the Einstein-Hilbert action:
\begin{align}
        S&=\frac{1}{16\pi G}\int_\mathcal{M} d^4x\,\sqrt{-g}R
        \\
        &=\frac{1}{16\pi G}\int_\mathbb{R} dr\int_{\Sigma_r} dtd\theta d\phi\, \mathcal{L},
\end{align}
where $\mathcal{L}=\mathcal{L}(r,q,q')$. Writing out the terms in the integrand for the time-like metric (\ref{metric}), we see
\begin{gather}
     \sqrt{-g}=Ne^{Y-\frac{X}{2}}\sin{\theta},
     \\
\begin{split}
     R=&2e^{X-Y}-\frac{N'X'}{N^3}+\frac{2N'Y'}{N^3}
     \\
     &-\frac{X'^2}{N^2}+\frac{2X'Y'}{N^2}-\frac{3Y'^2}{2N^2}+\frac{X''}{N^2}-\frac{2Y''}{N^2}.
\end{split}
\end{gather}
Integrating over the solid angle and integrating by parts, to eliminate second order terms, we reduce to a 2D first order action:
\begin{equation}
\begin{split}
    S=\int_{\mathbb{R}^2} drdt\,\frac{Ne^{Y-\frac{X}{2}}}{4G}&\left(2e^{X-Y}-\frac{X'^2}{2N^2}+\frac{Y'^2}{2N^2}\right)
    \\
    &+\text{Boundary Terms}.
\end{split}
\end{equation}
Discarding boundary terms, the remaining integrand is our symmetry-reduced Lagrangian. Through $p_q=\frac{\partial\mathcal{L}}{\partial q'}$, we calculate the conjugate momenta
\begin{align}
    p_X&=-\frac{e^{Y-\frac{X}{2}}X'}{4GN},
    \\
    p_Y&=\frac{e^{Y-\frac{X}{2}}Y'}{4GN},
\end{align}
and through the Legendre transform:
\begin{equation}
    \mathcal{H}=p_XX'+p_YY'-\mathcal{L},
\end{equation}
we derive the Hamiltonian for exterior Schwarzschild spacetime:
\begin{equation}
     \mathcal{H}=-2GNe^{\frac{X}{2}}\left[e^{-Y}(p_X^2-p_Y^2)+\frac{1}{4G^2}\right].
\end{equation}
For the interior spacetime, we simply repeat the calculation for the corresponding space-like 4D Ricci scalar. Alternatively there exists a duality transformation between the interior and exterior metrics:
\begin{align}
    N&\rightarrow-iN,
    \\
    X&\rightarrow X+i\pi,
    \\
    Y&\rightarrow Y+i\pi,
\end{align}
which translates to a duality between the Hamiltonians.

\section{Interfering WKB wavefunction}
\label{app:two}

A WKB analysis of both the exterior and interior spacetime, including ingoing and outgoing modes in equal mixture, produces a wavepacket over two trajectory branches, $t_\pm=X\pm Y_{\text{eff}}$:
\begin{widetext}
\begin{equation}
\begin{split}
    \Psi_{\text{e}\backslash\text{i}}=(-1)^{\frac{1}{4}}&\left(\frac{128\pi\sigma_m^2}{l_P^4m_0^2\pm e^Y}\right)^{\frac{1}{4}}\cdot
    \\
    &\cdot\left[\cosh{\left(\frac{P_{\text{e}\backslash\text{i}}(m_0)}{l_P^2}-\frac{m_0Y_{\text{eff}}^{\text{e}\backslash\text{i}}}{2}\right)}\Biggl\{e^{-\frac{t_+^2\sigma_m^2}{4}}\cos{\left(\frac{m_0t_+}{2}\right)}+e^{-\frac{t_-^2\sigma_m^2}{4}}\cos{\left(\frac{m_0t_-}{2}\right)}\Biggr\}\right.
    \\
    &\hspace{1em}-i\left.\sinh{\left(\frac{P_{\text{e}\backslash\text{i}}(m_0)}{l_P^2}-\frac{m_0Y_{\text{eff}}^{\text{e}\backslash\text{i}}}{2}\right)}\Biggl\{e^{-\frac{t_+^2\sigma_m^2}{4}}\sin{\left(\frac{m_0t_+}{2}\right)}-e^{-\frac{t_-^2\sigma_m^2}{4}}\sin{\left(\frac{m_0t_-}{2}\right)}\Biggr\}
    \right].
\end{split}
\end{equation}
\end{widetext}

\end{document}